\documentclass[epj]{svjour}     
\usepackage{graphicx,bbm,amsmath,amssymb,amsfonts}
\journalname{EPJD}
\usepackage{xcolor}
\newcommand{\gmz}{\gamma_{\hbox{\tiny{0}}}}
\newcommand{\gmu}{\gamma_{\hbox{\tiny{1}}}}
\newcommand{\ket}[1]{|#1\rangle}
\newcommand{\bra}[1]{ \langle\ #1 |}
\newcommand{\ketket}[1]{|#1\rangle\!\rangle}
\newcommand{\brabra}[1]{ \langle\!\langle #1 |}

\newcommand{\dketket}[1]{\left|\left.\! #1 \!\right\rangle\!\!\!\right\rangle}
\newcommand{\dbrabra}[1]{ \left\langle\!\!\!\left\langle \!#1 \right|\right.}
\newcommand{\cb}[1]{{\color{black} #1}}

\begin{document}
\title{Qubit systems subject to unbalanced random telegraph noise: 
quantum correlations, non-Markovianity and  teleportation}
\titlerunning{$ $ Qubits subject to unbalanced random telegraph noise}
\authorrunning{S. Daniotti {\it et al.} }
\author{Simone Daniotti, Claudia Benedetti, Matteo G. A. Paris}
\institute{Quantum Technology Lab, Dipartimento di Fisica {\em Aldo Pontremoli}, Universit\`a degli Studi di Milano, I-20133 Milano, Italy}
\date{\today}
%\maketitle
%
\abstract{
We address the dynamics of quantum correlations in a two-qubit
system subject to unbalanced random telegraph noise (RTN) and discuss
in details the similarities and the differences with the 
balanced case. We also evaluate quantum non-Marko\-vianity 
of the dynamical map. Finally, we discuss the effects of unbalanced 
RTN on teleportation, showing that noise imbalance may 
be exploited to mitigate decoherence and preserve
teleportation fidelity.}
\maketitle
\section{Introduction}
Quantum information processing based on solid-state qu\-bits 
\cite{ssq1,ssq2} plays a relevant role in current quantum technologies 
\cite{ssqn1}. Indeed, solid-state systems are scalable and highly 
controllable.  At the same time, these systems can be hardly isolated 
from their surroundings and thus the characterization of 
decoherence induced by the interaction with the environment 
is of primary importance when one is looking for practical 
implementations of quantum information processing 
\cite{w88,paladino14}.
\par
In  superconducting charge, phase or flux qubit, the 
computational basis $\{|0\rangle ,|1\rangle \}$ correspond 
to a fixed number of Cooper pairs, flux quanta, or charge 
oscillations in a Josephson junction, respectively 
\cite{makhlin01,Nakamura97,Nakamura99,mooij99,friedman00},
whereas proposals for solid-state qubits include quantum 
dot  and spin-based qubits in seminconductor nanostructures 
\cite{tarucha98,petta05,shi12}. In those implementations, the 
effects of phonons, electromagnetic and background charge 
fluctuations are relevant and definitely induce decoherence\cite{itakura03,hu06,gamble12}. 
\par
In particular, background charge fluctuations have been observed 
in different systems \cite{zorin96,krupenin98,jung04,stoop17}, 
e.g. linked to electrostatic potential fluctuations due  
to the dynamics of electrons trapped at impurity sites, which
are typically predominant at low frequencies \cite{kurdak97,paladino14}.
In these systems, fluctuations due to a single impurity leads 
to random telegraph noise (RTN), corresponding to a Lorentzian 
spectrum in the frequency domain. In turn, RTN was observed in 
many semiconductor devices, such as submicrometer 
metal-oxide-semiconductor field-effect transistors and  
me\-tal-insulator-metal tunnel junctions\cite{ralls83,kirton89,sung11,kaczer13}. 
The overall effect of several RTN sources have been suggested 
as the origin of the $1/f$ noise in electronic materials \cite{rt6}, 
as well as in any other context where the dynamics is governed 
by tunnelling \cite{rt7}.
\par
Motivated by the above considerations, and by 
 advances in  \cb{ implmentations of solid-state-based
quantum technologies, ranging from %environment engineering, 
     implementations
of quantum protocols  to 
eingeneering of quantum-classical interfaces \cite{reilly15,carlo09,steffen13,hirose16},
 we analyze how the introduction of 
imbalance in random telegraph noise affects the dephasing dynamics of 
two-qubit systems, for both the cases of local and global noises. 
This work allows us 
to make a thorough comparison with previous studies \cite{benedetti12,benedetti14}.
In particular, we are interested in the effect of unbalanced RTN on
resources for 
quantum technologies, such as 
entanglement and non-Markovianity \cite{wootters98,bylicka14}. 
Moreover, we study how the fidelity of the teleportation, 
 which exploits entanglement as a resource, is influenced by this kind of noise.
   }

%We address a simple system made of two qubits and
%assume that the environment may be modeled as a classical fluctuating 
%field, leading to pure dephasing. We analyze  the dynamics 
%of quantum correlations for independent and common environments, and 
%discuss in details the similarities and the differences between 
%the balanced and the unbalanced case.
We show that revivals of entanglement are present for certain 
values of the noise parameters, which we fully characterize.
We then evaluate the non-Markovianity of the dephasing map, by looking 
 first at the  single-qubit \cb{and  then at the two-qubits dynamics},   and we show 
 that memory effects are present and entaglement revivals live in the same
region of the parameter space.
In the final part of the paper, we discuss the effects of 
unbalanced RTN on the performances of teleportation
protocols, showing that noise imbalance may be exploited to mitigate 
decoherence and preserve teleportation fidelity. Overall, our results
suggest that even a modest engineering of RTN environments would represent
a resource for quantum information processing.
\par
The paper is structured as follows: in Section 
\ref{sec:model} we establish notation and 
introduce some preliminary concepts. %We describe  our interaction 
%model and how to introduce noise, and then briefly review 
%the quantification of entanglement, 
%non-Markovianity and teleportation fidelity. 
In Section \ref{sec:risenta}, we illustrate our 
results about the effects of unbalanced RTN on the 
dynamics of quantum correlations. Section 
\ref{sec:risnm} is devoted to non-Markovianity 
whereas in Section \ref{sec:ristel} we analyze 
teleportation fidelity in the presence of unbalanced 
RTN. Section \ref{sec:concl} closes the paper with 
some concluding remarks.
%%%%
\section{Preliminaries}
\label{sec:model}
\cb{In the three following subsections, we describe  our interaction 
model and how to introduce noise in it; we present the concepts of balanced and 
unbalanced RTN;  and  briefly review the 
quantification of entanglement, non-Markovianity, and teleportation fidelity. }
%In particular,
%in the five following subsections we describe our 
%interaction and noise models, and briefly review the 
%quantification  of entanglement, non-Marko\-vianity, 
%and teleportation fidelity. 
%%
\subsection{Two-qubit systems in a dephasing environment}
Throughout this paper, we mostly discuss the dynamics of 
two non-interacting qubits subject to a noisy environment 
which causes dephasing (a single qubit when evaluating non-Markovianity).
The interaction between the system 
and the environment can be either \textit{local}, i.e. the
two qubits are interacting with two indipendent environments
or \textit{global}, if a common environment affects both 
of them.
We assume that the noise corresponds to a fluctuating
field, which may be described as a stochastic 
process perturbing the energy splitting of the qubits.
Upon setting $\hbar = 1$, the two-qubit Hamiltonian may 
be written as
\begin{align}\label{hh}
H(t) = H_1(t) \otimes \mathbb{I}_2 + \mathbb{I}_1 \otimes H_2(t)
\end{align}
where $ H_{k}(t)$, $k=1,2$ is the single qubit Hamiltonian
\begin{align}\label{hk}
H_{k}(t) = \omega_0 \sigma_3 + \nu\, B_{k}(t) \sigma_3\,,
\end{align}
being $\omega_0$ the (equal) energy splitting of the qubits, $\nu$ 
a constant setting the amplitude of the system-environment coupling,
and $B_k(t)$ a stochastic process describing the fluctuating field.
The evolution of the global system is governed by the unitary 
$U(t) = U_1(t) \otimes U_1(t)$. Since $H_{k}$ commutes 
with itself at different times, we may write the evolution operator as:
\begin{align}
U_{k}(t) = \exp\left\{-i[\omega_0\, t + \nu\, \varphi_{k}(t)] \sigma_3\right\}
 \end{align}
where the noise-phase is given by
$
\varphi_{k}(t) = \int_0^t\!ds\, B_{k}(s)
$.
In the interaction picture (i.e. in a $\omega_0$-rotating 
frame for both  qubits) %the evolution operator is the diagonal matrix:  
% \begin{align}
%& U\! (t)  = \hbox{Diag}\! \left(e^{-i \nu \varphi_{\textsc{\tiny $+$} } (t)},
%	e^{-i \nu \varphi_{\textsc{\tiny $-$}}(t)},e^{i \nu \varphi_{\textsc{\tiny $-$}}(t)},
%	e^{i \nu \varphi_{\textsc{\tiny $+$} }(t)}  
%\right)\\
%& \varphi_{\textsc{\tiny $\pm$} } (t)  = \varphi_1(t) \pm \varphi_2(t).
%\end{align} 
the evolved state of the two-qubit system is thus 
given  by 
\begin{align}\label{rhot}
\rho(t) = \mathbb{E}_{\textsc{\tiny $B_1\!B_2$}}[U(t) \rho_0 U^\dagger(t)]\,,
\end{align}
where $\mathbb{E}_{\textsc{\tiny $B_1\!B_2$}}[...]$ denotes the ensemble average over 
all possible realizations of the stochastic processes $B_1(t)$, $B_2(t)$.
We assume that the two qubits are initially prepared in a generic 
Bell-state mixtures $\rho_0 = \sum_{k=0}^3 c_k\, 
\ketket{\frac{\sigma_k}{\sqrt{2}}}\brabra{\frac{\sigma_k}{\sqrt{2}}},$ where $\ketket{\frac{\sigma_k}{\sqrt{2}}}$, 
using the Choi-Jamiolkowski isomorphism 
$\ketket{\psi} = \sum_{jk} (\psi)_{jk}\, |j\rangle\otimes |k\rangle$,
denotes the $k$-th Bell state.
For example $\ketket{ \frac{\sigma_1}{\sqrt2}}=\cfrac{1}{\sqrt{2}}(\ket{01}+\ket{01})$. %and
%$\ketket {\frac{\sigma_3}{\sqrt2}}=\cfrac{1}{\sqrt{2}}(\ket{00}-\ket{11})$. 
It then follows that the single-realization density matrix 
$\rho^\prime = U(t) \rho_0 U^\dagger(t)$
is given by (we drop the time-dependency in order to simplify the notation)
\begin{align}\label{uevol}
\rho^\prime = \frac12  \everymath{\scriptstyle} 
 \begin{pmatrix}
  c_{03}^+ & 0 & 0 & c_{03}^-\, e^{-2 i \nu \varphi_{\textsc{\tiny +} } }\!\\
  0 & c_{12}^+ & c_{12}^-\, e^{-2 i \nu \varphi_-} & 0  \\
  0 & c_{12}^-\,e^{2 i \nu \varphi_-} & c_{12}^+ & 0 \\
 c_{03}^-\, e^{2 i \nu \varphi_+} & 0 & 0 &  c_{03}^+
  \end{pmatrix}\!
 \end{align}
where $c_{jk}^\pm = c_j\pm c_k$ and $\varphi_{\textsc{\tiny $\pm$} }  = \varphi_1 \pm \varphi_2$.
The explicit evaluation of the output density matrix 
$\rho(t)$ in Eq. (\ref{rhot}) depends on the specific 
features of the noise model. 
In particular, for the interaction Hamiltonian of Eq. (\ref{hh})
the relevant quantities are the { time-dependent} averages 
$\mathbb{E}_{\textsc{\tiny $B_1\!B_2$}}\!\!\left[e^{\pm 2 i \nu \varphi_{\textsc{\tiny $\pm$} } }\right]$
over the realizations of the stochastic processes 
describing the external fields. In the following, 
we will discuss in details two scenarios: the case 
of identical but independent environments (IE) for the two quits,
the  fields $B_1(t)$ and $B_2(t)$ are independent but they are the same identical process   $B(t)$,
 and that of a common environment 
(CE), corresponding to $B_1(t)= B_2(t) = B(t)$. 
In the case of independent environments, we have:
\begin{align}
\mathbb{E}_{\textsc{\tiny $B_1\!B_2$}}\left[e^{\pm 2 i \nu \left[\varphi_1(t) 
+ \varphi_2(t)\right]}\right] &= \mathbb{E}_{B}\left[e^{\pm 2 i \nu \varphi(t)}\right]^2 \label{iel} \\
\mathbb{E}_{\textsc{\tiny $B_1\!B_2$}}\left[e^{\pm 2 i \nu \left[\varphi_1(t) 
- \varphi_2(t)\right]}\right] &= \mathbb{E}_{B}\left[e^{\pm
 2 i \nu \varphi(t)}\right] \mathbb{E}_{B}\left[e^{\mp 2 i \nu \varphi(t)}\right]\,, \notag
\end{align} 
while in the case of a common environment
\begin{align}
\mathbb{E}_{\textsc{\tiny $B_1\!B_2$}}\left[e^{\pm 2 i \nu \left[\varphi_1(t) 
+ \varphi_2(t)\right]}\right] &= \mathbb{E}_{B}\left[e^{\pm 4 i \nu \varphi(t)}\right]  \label{cel} \\
\mathbb{E}_{\textsc{\tiny $B_1\!B_2$}}\left[e^{\pm 2 i \nu \left[\varphi_1(t) 
- \varphi_2(t)\right]}\right] &= 1.
\end{align}
Eqs. \eqref{iel} is due to the fact that for identical  independent processes,
$
\mathbb{E}_{\textsc{\tiny $B_1\!B_2$}}\!\!\left[e^{\pm 2 i \nu (\varphi_1 
+ \varphi_2) }\right] \!=$$\mathbb{E}_{\textsc{\tiny $B_1$}}\!\left[e^{\pm 2 i \nu \varphi_1} \right]$$
\mathbb{E}_{\textsc{\tiny $B_2$}}\!\left[e^{\pm 2 i \nu \varphi_2}\right]$ $=\mathbb{E}_{\textsc{\tiny $B$}}\!\left[e^{\pm 2 i \nu \varphi}\right]^{\!2}
$.
All the above quantities themselves correspond to the characteristic function
of the stochastic processes describing the noise in the different cases.
%%%%%%%%%%%%%%%%%%%%%%%%%%%%%%%%%%%%%%%
\subsection{{ Balanced and unbalanced random telegraph noise}}
%In physical systems described by Hamiltonians of the form 
%(\ref{hk}) the noise due to the interaction with the environment
%induces fluctuations in the energy splitting between the two 
%levels of the qubits. 
%Indeed, this kind of model is suitable to describe the bistable fluctuations that arise in many nano-devices based 
%on semiconductors, metals, and superconductors that are due, for example, to defects that act as traps for charged particles.
\cb{The terms $B_k(t)$ in Eq. \eqref{hk}  describe classical fluctuations. 
In our model, we  describe these fluctuations as a random telegraph noise, }
 which consists of random switching
between an up and a down state  at given rates $\tilde\gamma_k$,  $k=0,1$ and that may affect quantities like a current or a voltage.
If the two rates are equal, i.e. the probability of switching 
from the up to down state and viceversa are the same, we speak of 
{\em balanced} random telegraph noise (BRT), whereas the opposite 
case is referred to as {\em unbalanced} random telegraph noise 
(URT) \cite{urt1,urt2}. Asymmetry may be due, for example, to 
the difference between the Fermi energy of the electron
reservoir and the energy level of the impurity sites.
 Both  cases of balanced and unbalanced RTN have been experimentally observed  \cite{buehler04,fuji00,kamioka14}.
\par
RTN corresponds to a non-Gaussian stochastic 
processes with a Lorentzian spectrum. Remarkably, for both the 
balanced and the unbalanced case, the characteristic functions 
of Eqs. (\ref{iel}) and (\ref{cel}) are amenable to an analytic evaluation \cite{urt2,bergli06,bergli09,fern17}.
 In terms of the rescaled rates $\gamma_k = 
\tilde\gamma_k/\nu$ and time $\tau = \nu t$, we have:
\begin{align}
\!\!\!\!\Lambda_n^b\!(\tau,\gamma)\! &=\!\mathbb{E}_{\hbox{\tiny{B\!R\!T}}}\!\!\left[e^{i n \varphi(\tau)}\!\right] \!= \! e^{-\gamma \tau}\!\!\left[\!\cosh \delta_b \tau + \frac{\gamma}{\delta_b}\sinh \delta_b \tau\! \right]  \label{lbrt}\\
\delta_b^2 & = \gamma^2-n^2\,,\label{brt}
\end{align}
for the balanced case with $\gmz=\gmu=\gamma$ 
and 
\begin{align}
\!\!\!\!\!\!\!\Lambda_n^u(\!\tau,\!\gmz,\!\gmu\!)\! &=\!\mathbb{E}_{\hbox{\tiny{U\!R\!T}}}\!\!\left[\!e^{\!i n \varphi(\tau)}\!\right] 
\! = \!  e^{-\overline\gamma\! \tau}\!\!\left[\!\cosh \delta_u \tau \!+\! \frac{\overline\gamma}{\delta_u}\!\sinh \delta_u \tau\! \right]  \label{lurt}\\
\delta_u^2 & = {\overline\gamma}^2-n^2+ 2 i \,n  \epsilon \,,\label{urt} \\
\overline\gamma & = \frac12 (\gmz+\gmu) \quad \epsilon = \frac12 (\gmz-\gmu) \notag 
\end{align}
for the unbalanced case. In both cases, \cb{ $n$ is a real number and } we have assumed 
that the process starts with  equal probability in one of the two 
possible values. As it is apparent from Eqs. (\ref{brt}) and (\ref{urt}) the behaviour of the characteristic functions may be either monotone or oscillatory in time, depending on the values of the switching rates.
\begin{table}[t!]
\center{
\begin{tabular}{|c|c|c|c|c|}
\hline$\mathbb{E}_{B_1B_2}$[...] & CE URT& CE BRT & IE URT & IE BRT \\
\hline $e^{-2 i (\varphi_1+ \varphi_2)}$ & $\Lambda_{-4}^u$ &$\Lambda_{4}^b$ & $\Big( \Lambda_{-2}^u\Big)^2$& $\Big( \Lambda_2^b\Big)^2$ \\
\hline $e^{-2 i (\varphi_1- \varphi_2)}$ & $1$ & $1$ &$\Lambda_2^u \Lambda_{-2}^u$ & $\Big( \Lambda_2^b\Big)^2$\\
\hline $e^{+2 i (\varphi_1- \varphi_2)}$ & $1$ & $1$ &$\Lambda_2^u \Lambda_{-2}^u$ & $\Big( \Lambda_2^b\Big)^2$\\
\hline $e^{+2 i (\varphi_1+ \varphi_2)}$ & $\Lambda_{4}^u$ &
$\Lambda_{4}^b$ & $\Big( \Lambda_2^u\Big)^2$ & $\Big( \Lambda_2^b\Big)^2$\\
\hline
\end{tabular}}
\caption{The process averages needed to evaluate the evolved 
density matrix $\rho(t)$ of Eq. (\ref{rhot}), for the different 
kinds of noise considered in this paper. The functions 
$\Lambda_n^k\equiv\Lambda_n^k(\tau)$ for \cb{$k=b,u$} are those
given in \cb{ Eqs. (\ref{lbrt}) and (\ref{lurt})}.}
\label{tab1}
\end{table}
\par\noindent
We also notice that in the balanced case we 
have $\Lambda_n^b(\tau,\gamma) = \Lambda_{-n}^b(\tau,\gamma)$, whereas in 
the unbalanced case we need to exchange the role of the two rates 
$\Lambda_n^u(\tau,\gmz,\gmu) = \Lambda_{-n}^u(\tau,\gmu,\gmz)$ in order to have the same symmetry. We also notice that $\Lambda^u_{-n}(\tau,\gmz,\gmu)=\left[\Lambda^{u }_{n}(\tau,\gmz,\gmu)\right]^*$, where $*$ stands for conjugation, and thus
$| \Lambda^u_{-n}(\tau,\gmz,\gmu) |=| \Lambda^u_{n}(\tau,\gmz,\gmu) |$. 
Overall, taking into account the symmetries of the characteristic
functions and the nature (independent or common) of the noise, the
process averages of the phase factors in Eq. (\ref{uevol}) may be 
summarised in the Table \ref{tab1} where, for the sake of simplicity, 
we are omitting the explicit dependence on time and on the rates.
%%%%%%%%%%%%%%%%%%%%%%%%%%%%%%%%%%%%%%%
\subsection{Quantification of entanglement, non-Markovianity and teleportation fidelity}
In two-qubit systems, entanglement may be quantified by several
measures \cite{horod09,toth09,plenio07}. Among the possible
entanglement monotones, we focus on negativity 
\cite{vidal02}, defined as
\begin{equation}
N_\rho=||\rho^{\hbox{\tiny pt}}||_1-1 = 2\left|\sum_k \lambda_k^-\right|\,,
\label{neg1}
\end{equation}
where $||\cdot||$ is the trace norm $||A||_1= 
\text{Tr}\left[\sqrt{A^{\dagger}A}\right]$
and $\rho^{\hbox{\tiny pt}}$ denotes the partial transpose of 
the density matrix with respect to one of the subsystems.
In other words, the negativity is absolute value of the 
sum of the negative eigenvalues $\lambda^-$ of $\rho^{\hbox{\tiny pt}}$.
Notice that the above definition slightly differ from the 
original one \cite{vidal02} in order to bound the negativity 
between 0 (for separable states) and 1 (for maximally 
entangled states). For the dephasing channels 
arising from BRT, the dynamics of negativity has been 
studied, for a system initially prepared in a Bell state
\cite{benedetti12}. 
%More generally, for the system initially 
%prepared in a mixture of Bell states, the entanglement negativity 
%$N(\tau)$ at time $\tau$ between two qubits subject to BRT noise takes 
%the form:
%\begin{align}
% \everymath{\scriptstyle} 
%N(\tau)& =  \frac12 \Big[ |\ct{1} + \ct{2}+ \Lambda_+ (\ct{0} - \ct{3})| \notag \\ 
%& + |\ct{0} + \Lambda_- (\ct{1} - \ct{2}) + \ct{3}|+  
%  |\ct{0} - \Lambda_- (\ct{1} - \ct{2}) + \ct{3}| \notag \\ & +
%| \ct{1} +\ct{2} - \Lambda_+(\ct{0}-\ct{3})|
%-2\Big]
%\end{align}
%where the coefficients $\Lambda_{\pm}=$ are given in Table \ref{tab1} 
%for independent and common BRT, respectively.
\par
\cb{
Together with entanglement, Non-Markovianity can also be considered a resource for quantum information processing tasks \cite{bylicka14,laine14,dong18}. Quantum non-Markovianity can be quantified which in terms of backflow of 
information from the environment to the system.
The idea is that Markovian dynamics tends to reduce the distinguishability 
between two initial states while non-Markovianity is linked 
with a regrowth in distinguishability \cite{blp}. Indeed a non-monotonic behavior in the
trace distance between properly optimized initial states $\rho_1(0)$ and $\rho_2(0)$ is a signature of memory effects. Non-Markovianity can thus be defined as:
\begin{align}
\mathcal{N}\!(\tau) = \underset{\rho_1(0)\rho_2(0)}{\max} \,\underset{ \dot{D} > 0}{\int_0 ^{\tau}}\dot{D}[\rho_1(t'),\rho_2(t')] dt' .
\end{align}
where $D[\rho_1,\rho_2]=\frac12 ||\rho_1-\rho_2||_1$ is the trace distance and $\dot{D}$ indicates its derivative with respect to time (see appendix \ref{app1} for more details).
In general, the optimization over the initial pair of states  is difficult 
to compute. However, for single-qubit dephasing the optimal trace distance is known  \cite{he11}: 
\begin{align} 
D(\tau) = 
\underset{
\rho_{\text{\tiny 1}(0)}
\rho_{\text{\tiny 2}(0)}
}{\max}  D[ \rho_{\text{\tiny1}} (\tau),  \rho_{\text{\tiny 2}} (\tau)]
= \left|\Lambda(\tau)\right|\,, 
\label{otd}
\end{align}
which is referred to as the optimal trace distance.
 In the following, we will 
analyze the behavior of $\mathcal{N}(\tau)$ as a function of time and of the 
switching rates, thus generalizing to URT the study that has been 
done for BRT \cite{benedetti14}.
\par
Quantum teleportation is an example of  a protocol that exploits quantum correlations as a resource in order to teleport an unknown
quantum state between two distant locations. Details of the protocols are found in appendix \ref{app2}.
In realistic situations, where noise corrupts 
entanglement, the teleported state $\rho_\psi$ may 
not be equal to the input state $\ket{\psi}$. Let us denote by 
$\rho_\psi \equiv {\cal E} [|\psi\rangle\langle\psi|]$
the state at Bob's site, where ${\cal E}$ is a quantum
operation  describing the
the overall action of the {\it imperfect} teleportation scheme. 
In this case, a convenient figure of merit to globally assess the 
protocol is the average fidelity, i.e. the input-output fidelity
averaged over all possible initial states of  qubit to be teleported:
\begin{align} \label{avF}
F_{av} &=\frac{1}{4\pi} \int_0^{\pi}\!\!\!d\phi\int_0^{2\pi}\!\!\!d\theta\,\sin\!\theta\, \langle\psi | \rho_\psi | \psi\rangle\,.
\end{align}
}

%%%%%%%%%%%%%%
%%%%%%%%%%%%%%%%%%%%%%%%%%%%%%
\section{Dynamics of entanglement}\label{sec:risenta}
Let us consider a two-qubit system initially prepared 
in a mixture of Bell states and then subject to unbalanced 
random telegraph noise.
The evaluation of the negativity  may be done 
analytically but the expression is cumbersome, and will 
not be reported here. When the initial state is a Bell state $\ketket{\frac{\sigma_k}{\sqrt{2}}}$ 
with $k=0,1,2,3$, negativity is given by
\begin{equation}
\begin{aligned}
&N_k^{\text{\tiny IE}}(\tau)=\frac12\left(| \Lambda^u_{-2}|^2 +| \Lambda^u_{2}|^2\right)=| \Lambda^u_{2}|^2&\text{  for }k=0,3\label{nie}\\
&N_k^{\text{\tiny IE}}(\tau)=1&\text{  for }k=1,2
\end{aligned}
\end{equation}
in the case of independent environments, and by
\begin{equation}
\begin{aligned}
&N_k^{\text{\tiny CE}}(\tau)=\frac12\left(| \Lambda^u_{-4}| +| \Lambda^u_{4}|\right)=| \Lambda^u_{4}|&\text{  for }k=0,3\label{nce}\\ 
&N_k^{\text{\tiny CE}}(\tau)=1&\text{  for }k=1,2
\end{aligned}
\end{equation}
for a common environment.
The Bell states $\ketket{\frac{\sigma_1}{\sqrt{2}}}$ and $\ketket{\frac{\sigma_2}{\sqrt{2}}}$ 
live in a decoeherence-free subspace and are not affected by 
decoherence. For this reason, we focus henceforth on 
the state $\ketket{\frac{\sigma_0}{\sqrt{2}}}$.
Note that Eqs. \eqref{nie} and \eqref{nce} for $\gmz=\gmu=\gamma$, coincide with the 
expressions for the negativity in the BRT case \cite{benedetti12}. Moreover, 
\begin{align}
N(\tau,\gmz,\gmu)=N(\tau,\gmu,\gmz)\,, \quad
N(\tau,\gmz,\gmz+\delta) \stackrel{\delta \gg 1}{\simeq}1,
\end{align}
i.e. negativity is invariant under the exchange of the switching rates
and it remains constant if the difference between the switching rates 
is large. 
\begin{figure}[t!]
\centering
\includegraphics[width=0.8\columnwidth]{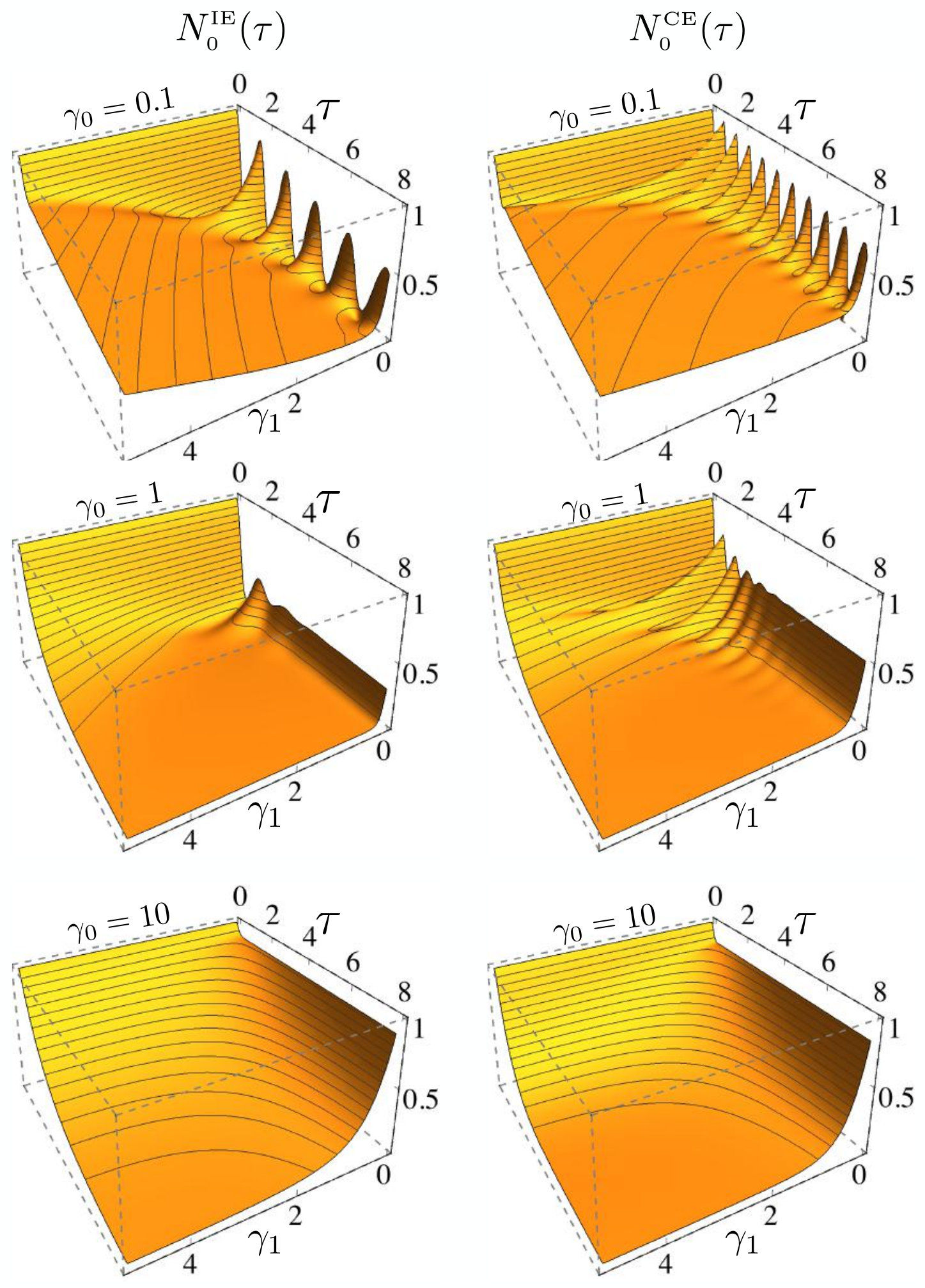}
\caption{Dynamics of entanglement of the two qubits subject 
to independent (left column) and a common (right column) URT 
as a function of the dimensionless time $\tau$ and the 
switching rate $\gmu$, for different values of the switching 
rate $\gmz$. From top to bottom we have $\gmz=0.1,1,10$ in both
columns.}
\label{enta}
\end{figure}
%%%
\par
\begin{figure}[t!]
\centering
\includegraphics[width=0.8\columnwidth]{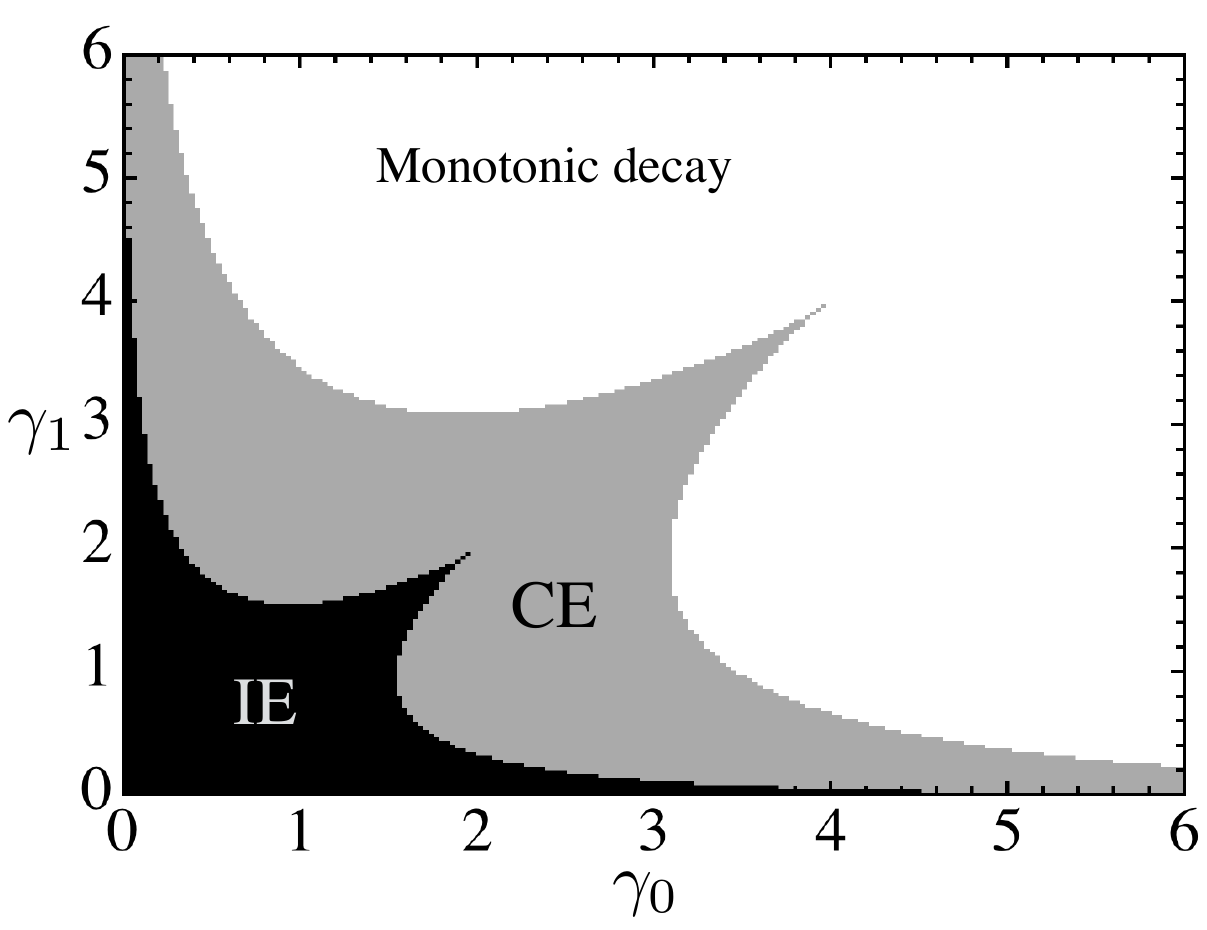}
\caption{Revivals of entanglement. The shaded areas represent the regions 
of the parameter space $\{\gmz,\gmu\}$ in which one observes 
oscillation in the entanglement dynamics. The black area is for 
independent environmentsm and the gray one for common environments. 
The cusps correspond to $\gmz=\gmu=2$ for independent environments, 
and to $\gmz=\gmu=4$ for common environments.}
\label{enta2}
\end{figure}

In Fig. \ref{enta} we show the time evolution of the negativity 
as a function of the dimensionless time $\tau$ for different values of the 
switching rates $\gmz$ and $\gmu$ for both independent and common 
URT. A large variety of behaviors emerge: there exists values of the 
switching rates for which the negativity evolves monotonically in time, 
while for others it display  oscillations. In addition, there are regimes 
in which it decays to zero and other in which it saturates to a certain value.
Typically, for small values of the switching rates revivals of quantum 
correlations are present. On the contrary, very large values of $\gmz$ and 
$\gmu$ lead to a monotonic evolution in time. In the intermediate regime, 
where one switching rate is large and the other small, revivals are present 
if the switching rates belong the a specific region of the parameter space 
$\{\gmz,\gmu\}$, as shown in Fig. \ref{enta2}.
When revivals are present, the effect of a common environment is to double 
the frequency of the oscillations and to increase their height compared 
to independent URT, thus leading to stronger quantum correlations. On the 
contrary, in case of a monotonic behavior, the effect of a common
environment is to lead to a faster loss of correlations.
The cusps in Fig. \ref{enta2} correspond to $\gmz=\gmu=2$ for independent 
environments, and to $\gmz=\gmu=4$ for common environments.\\
\cb{In order to complete our analysis, we  also consider a third scenario, where only one
qubit is subject to URT, e.g. $B_1(t)=0$ in eq.\eqref{hk}. 
We suppose  the the qubit pair is initially in a Bell state. The dynamics of negativity
for all four Bell states then reads:
\begin{align}
N^{\text{\tiny 1q}}(\tau)=\frac12\left(| \Lambda^u_{-2}| +| \Lambda^u_{2}|\right)=| \Lambda^u_{2}|.\label{n1q}
\end{align}
}
%%%%%%%%%%%%%%%%%%%%%%%%
%%%%%%%%%%%%%%%%%%%%%%%%%%%%%%%%%%
\section{Non-Markovianity of URT dephasing}\label{sec:risnm}
\begin{figure}[t!]
\centering
\includegraphics[width=0.9\columnwidth]{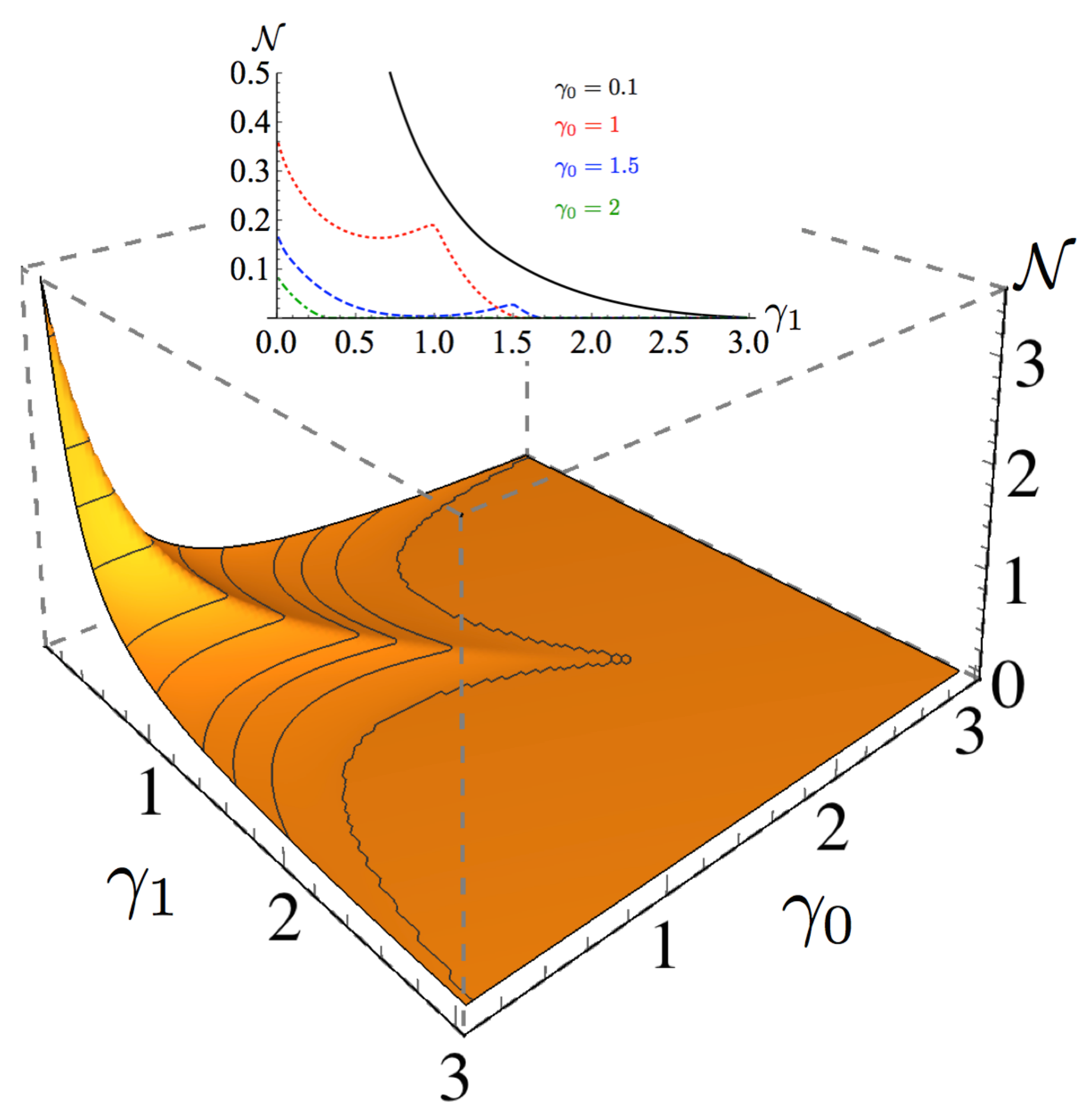}
\caption{Main panel: BLP measure of Non-Markovianity as a function of 
$\gmz$ and $\gmu$ calculated at time $\tau=10$. The inset shows 
slices of the 3D plot for fixed values of $\gmz$, i.e. the behavior 
of non-Markovianity as a function of $\gmu$ for different fixed 
values of $\gmz$.}
\label{fig_nm}
\end{figure}

As mentioned in Section \ref{sec_nmth},  the optimal trace distance of the
dephasing map is given by the absolute value of the decoherence factor, 
see Eq. \eqref{otd}.
\cb{ Moreover, we numerically investigate the  optimal pair in the case of two qubits
subject to local independent environments, and we found that the initial states $\ketket{\!+\!+}$ and $\ketket{\!-\!-}$  
yield the maximum trace distance. A trivial calculation shows that 
the corresponding trace distance is equal to the absolute value of the decoherence factor \eqref{otd}.
This means that the single-qubit and two-qubit non-Markovianity coincides
 and the
}
optimal trace distance  reads:
\begin{align}
D(\tau)&=|\Lambda_{-2}^a(\gmz,\gmu,\tau)|=|\Lambda_{2}^a(\gmz,\gmu,\tau)|,
\end{align}
with the corresponding information flow:
\begin{align}
\dot{D}[\tau]=\Lambda^a_2(\gmz,\gmu,\tau)\,\frac{d}{dt}\Lambda^{a}_{-2}(\gmz,\gmu,\tau)+c.c.
\end{align}
where $c.c.$ stands for complex congiugate.
Inserting this expression into Eq. \eqref{nonmar}, we obtain the expression 
for BLP non-Markovianity.
We report the behavior of $\mathcal{N}(\tau)$ at time $\tau=10$ in Fig. \ref{fig_nm} 
as a function of $\gmz$ and $\gmu$. In order to understand the 
behavior  of the non-Markovianity, we first recall that BLP measure of 
non-Markovianity is different from zero only if there are revivals in 
the optimal trace distance.  \cb{The expression of the optimal 
trace distance in Eq.\eqref{otd}
exactly coincides with the entanglement, quantified by the concurrence, between a single 
qubit and an ancilla system  \cite{liu11}.
}Moreover, we notice that we have already 
analyzed a similar expression in studying entanglement between two 
interacting qubits, see Eq.\eqref{nie}. This means that the regions 
of the parameter space for which entanglement has revivals coincides 
exactly with the region where the optimal trace distance has revivals, 
i.e. BLP measure is non-zero. We may now easily illustrate 
the the behavior shown in the main panel of Fig. \ref {fig_nm}: 
the BLP measure is different from zero for the values 
$\{\gmz,\gmu\}$ that lies inside the black area in Fig. \ref{enta2}. 
This may be seen also in the inset, which shows 
slices of the 3D plot for fixed values of $\gmz$, i.e. the behavior 
of non-Markovianity as a function of $\gmu$ for different fixed 
values of $\gmz$.
%%%
For BRT noise \cite{benedetti14}, there is a threshold at $\gamma=2$, separating 
Markovian and non-Markovian dynamics. In the present case of URT a more complex structure arises, and the threshold for backflow of information changes according 
to fig. \ref{fig_nm}. We notice also that the balanced case 
coincides with the cusp at $\gmz=\gmu$ and, in general, do not coincide with 
the largests values of non-Markovianity, i.e. imbalance leads to stronger memory effects.  From the inset of Fig. \ref{fig_nm} we also see that a general feature 
of NM is that the smaller the values of the switching rates, the larger the
value of NM. Already when one of the switching rate starts to increase, NM quickly 
vanishes, confirming the idea that slow noise is connected to non-Markovian dynamics.
%%%%%%%%%%%%%%%%%%%%%%%%%%%%%%%%%%%%%%
\section{Noisy quantum teleportation}\label{sec:ristel}
Here we analyze teleportation fidelity when the Bob's qubit (qubit 3 in 
Fig. \ref{qtel}) is subject to URT.
\cb{ This scenario corresponds to the situation where Alice generates the entangled pair
and then she sends one qubit to Bob through a channel that is affected by noise. }
The the initial Bell state $\ketket{\frac{\sigma_0}{\sqrt{2}}}$ is  
subject to the decoherent evolution given by
\begin{align}
R=(\mathbb{I} \otimes \mathcal{E}) \left[\dketket{\frac{\sigma_0}{\sqrt{2}}}\!\dbrabra{\frac{\sigma_0}{\sqrt{2}}}\right],
\label{evo0}
\end{align}
where $\mathcal{E}$ is the quantum map describing the URT dephasing, \cb{which induces the 
dynamical evolution of entanglement given by Eq.\eqref{n1q}.}
The map \eqref{evo0} can be expressed in terms of Kraus operators as:
\begin{align}
\mathcal{E}[\rho_0]=\sum_{k=1}^2M_k(\tau)\rho_0M_k^{\dagger}(\tau)\label{krausrep}
\end{align}
with 
\begin{align}
M_1(\tau)&=\sqrt{ \frac{1-|\Lambda(\tau)|}{2}}
\left(
\begin{array}{ll}
-\frac{\Lambda (\tau)}{|\Lambda (\tau)|}&0\\
0&1
\end{array} \right)%=\frac{\sqrt{1-|\Lambda\!(\tau)|}}{\sqrt{2}}\, \mu_1(\tau)\\
\nonumber\\
\nonumber\\
M_2(\tau)&=\sqrt{ \frac{1+|\Lambda(\tau)|}{2}}
\left(
\begin{array}{ll}
\frac{\Lambda(\tau)}{|\Lambda(\tau)|}&0\\
0&1
\end{array} \right)%= \frac{\sqrt{1+|\Lambda(\tau)|}}{\sqrt{2}}\, \mu_2(\tau)
\end{align}
and $\Lambda(\tau)=\Lambda^a_2(\tau,\gmz,\gmu)$ is the URT dephasing factor of 
Eq. \eqref{urt}. 
In the case of balanced RTN, Eq. \eqref{krausrep} reduces to the more familiar expression:
\begin{equation}
\mathcal{E}^s[\rho_0]=\frac{1-\Lambda(\tau)}{2}\,\sigma_3\,\rho_0\sigma_3+\frac{1+\Lambda(\tau)}{2}\rho_0, 
\end{equation}
with $\Lambda(\tau)=\Lambda^s_2(\tau,\gamma)$  
the dephasing factor in Eq. \eqref{brt}.
By explicitly calculating the evolution in \eqref{evo0}, we obtain: 
\begin{align}
 R(\tau) = \frac12 \Big( \dketket{M_1(\tau)\,}\dbrabra{\,M_1(\tau)} +
 \dketket{M_2(\tau)\,}\dbrabra{\,M_2(\tau)} \Big)
\end{align}
where we used the property $O_1\otimes O_2 \ketket{\psi}=\ketket{O_1\psi \,O_2^T}$. 
After straightforward calculations \cite{paris12}, 
we obtain  Bob's conditional state
\begin{align}
\rho_\psi(\tau)=\sum_{j=1}^2\sigma_k\,M_j(\tau)\,\sigma_k\,P_\psi\,\sigma_k\,M_j^*(\tau)\sigma_k\,.
\end{align}
When $\Lambda(\tau)=1$, we recover the noiseless case which allows one to perfectly teleport the initial state. In the most general case, the input-output 
fidelity is given by
\begin{align}
F(\tau)=
\frac12 \Big\{ 1+\mathcal{R}[\Lambda (\tau)]+\big(1-\mathcal{R}[\Lambda (\tau)]\cos^2\theta\big)  \Big\}
\end{align}
where $\mathcal{R}[x]$ stands for the real part of $x$. 
The corresponding average fidelity, see Eq. (\ref{avF}),  
reads as follows
\begin{align}
F_{av}(\tau) =
\frac{1}{3} \Big(2+\mathcal{R}[\Lambda (\tau)]\Big)\,.\label{tel1q}
\end{align}
\begin{figure}[t!]
\centering
\includegraphics[width=0.8\columnwidth]{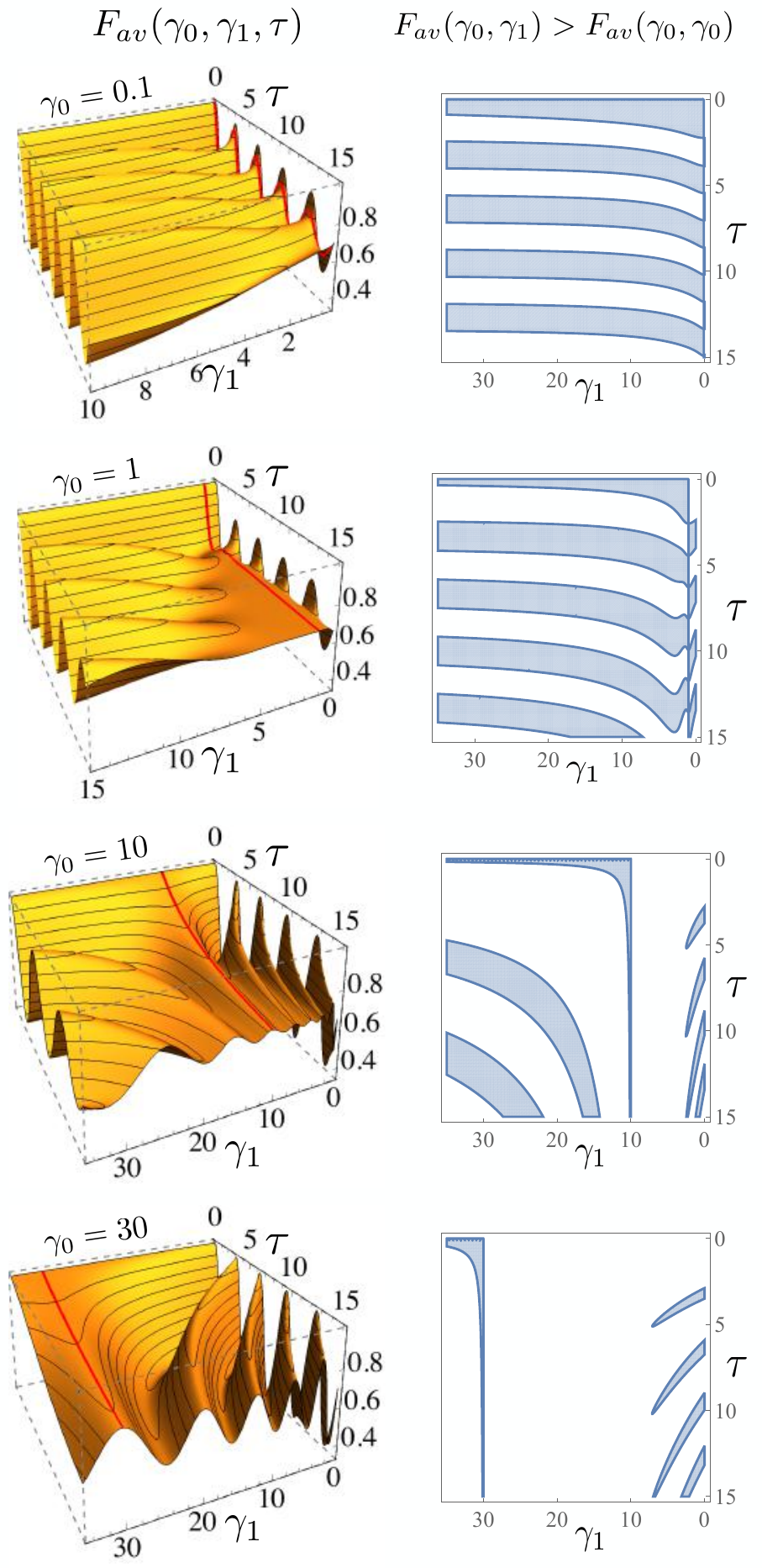}
\caption{Left column: Average fidelity as a function of the dimensionless 
time $\tau$ and the switching rate $\gmu$, for different values of 
$\gmz=0.1,\,1,\,10,\,30$ from top to bottom. The red lines are guides for 
the eye highlighting the case $F_{av}(\gmz, \gmz,\tau)$.
Right column: Area in the $\{\gmu,\tau\}$ parameter-space where  
$F_{av}(\gmz,\gmu,\tau)>F_{av}(\gmz,\gmz,\tau)$, for the same values of $\gmz$ represented in the left column ($\gmz=0.1,\,1,\,10,\,30$ from top to bottom). }
\label{fig_fide}
\end{figure}
\par\noindent
The behavior of the average fidelity is shown in Fig. \ref{fig_fide}. On the left column we show $F_{av}$ as  a function of the dimensionless time and the switching 
rate $\gmu$, for different values of $\gmz$. Different temporal behaviors arise, depending on the values the switching rates. Indeed, as a function of time, it is possible to find either non-monotonic $F_{av}$ or monotonic decaying average fidelity.
Although it is not trivial to describe the different regimes for the fidelity 
in general, some features emerge. Oscillations are present either when the two 
values of the switching rates are small or when one is small and the other large. 
In the last case, oscillations usually achieve a larger amplitude. This has a clear physical interpretation: if one can tune the length of the Bob's noisy channel 
such that it corresponds to a maximum in the fidelity oscillation, it is possible 
to teleport the inital state with fidelity  almost equal to one. 
Another feature emerging from the plot is that $F_{av}(\gmz,\gmz,\tau)$ 
has a monotonic behavior for any $\gmz>2$. However, even  a small unbalance 
between the two switching rates may again produce revivals in the fidelity 
$F_{av}(\gmz,\gmu,\tau)$ with $\gmz,\gmu>2$.
\par
In the right column of Fig. \ref{fig_fide}, we compare the value of the 
average fidelity in the case of URT with the case of BRT. We notice that 
there exist values of the parameters for which the fidelity in the case of 
URT is larger than that obtained in the presence of BRT, meaning that the 
quantum teleportation performances can be improved by unbalancing the 
switching rates.  Moreover, as it is apparent from the plots, when the 
switching rates are small, it is very easy to outperform fidelity of the 
balanced case (see for example the first plot on the right column). On the 
contrary, the region of the parameter space that allows one to exceed 
the BRT fidelity for large rates is very small (see bottom plot on the 
right column) and it becomes difficult to improve the performances of 
the teleportation protocol in the case of BRT with  large switching rates.
\cb{Analogous considerations can be made if we modify the teleportation
protocol to add noise on both parties, i.e. both Alices's and Bob's qubit are subject 
to local and independent URT. In this case, the Hamiltonian is described by Eqs. \eqref{hh} and \eqref{hk} 
with $B_1(t)\neq B_2(t)$ and a
simple calculation leads to an average fidelity:
\begin{align}
F_{av}(\tau)=\frac{1}{3}\left(2+\mathcal{R}\left[\Lambda^2(\tau)\right]\right).\label{tel2q}
\end{align} 
Teleportation is based upon the presence of entanglement, however the expression
of the average fidelity  is not directly related to the negativity , since the first depends on the
real part of  $\Lambda(\tau)$ or $\Lambda^2(\tau)$ (cfr Eqs. \eqref{tel1q} and \eqref{tel2q}), while entanglement varies as their absolute value (see Eqs. \eqref{n1q} and \eqref{nie}).  In the case of balanced RTN revivals in the fidelity and entanglement coincides , since $\Lambda(\tau)$ 
is a real quantity, but in the most general case of URT this is not true, and there is not a simple connection between the temporal behaviors of the two quantities. 
  }
%%%%%%%%%%%%%%%%%%%%%%%%
\section{Conclusions}\label{sec:concl}
We have addressed the dynamics of quantum correlations in a two-qubit
system subject to unbalanced random telegraph noise and have discussed
in details the similarities and the differences with the balanced case.
In particular, we have analyzed the effect of URT noise on entanglement, 
non-Markovianity  and
teleportation fidelity, and have individuated different working regimes.
\par
We have found that entanglement of an initial Bell pair subject to either 
independent or common environments shows revivals as a function of time in 
a specific region of the  $\{\gmz,\gmu\}$ parameter space.  A common environment 
leads to faster oscillations and  to a stronger non-monotonic behavior, i.e a 
larger region in the parameter space leading to oscillations. 
We have linked revivals of entanglement (with independent environments) to 
non-Markovianity of the \cb{two-qubit map}. This is due to the fact that both 
quantities depend on the absolute values of the decoherence factor. 
Finally, we have addressed fidelity of teleportation protocol for the shared 
Bell state subject to (one-side) URT dephasing and found that fidelity depends 
on the real part of the decoherence factor. A variety of different behaviors 
arose, ranging from a monotonic-decaying to the presence of revivals, with unbalanced
noise that allows one to increase fidelity with respect to the balanced case.
\par
Overall, our results show that noise imbalance may be exploited to 
mitigate decoherence and preserve teleportation fidelity, thus 
suggesting that even a modest engineering of environment, 
in the direction of making the switching rates different, permits 
to better preserve quantum features and, in turn, to improve 
performances of quantum information protocols in the presence of noise.
%%%%%%%%%%%%%%%%%%%
\section*{Acknowledgment}
This work has been supported by SERB through project 
VJR/2017/000011. MGAP is member of GNFM-INdAM. 
\section*{Author contribution statement}
SD, CB and MGAP contributed to the design and implementation
of the research, to the analysis of the results and to the
writing of the manuscript.

%%%%
\appendix

\section{Appendix}\label{app}
\cb{ In this appendix we review the definitions of the non-Markovianity in terms of information backflow and we review the teleportation protocol. }
\subsection{Quantification of (quantum) non-Markovianity}\label{app1}
\label{sec_nmth}
In classical physics, a Markovian process usually refers to 
a stochastic model where the probability distribution of 
the events depends only on the state attained in the 
previous event. Non-Markovianity (NM) is accordingly defined
as the violation of this conditions, and usually involves
the appearance of memory effects in the dynamics of the
considered systems. When one moves to open quantum systems, 
one realises that memory effects play an important role 
and, together with entanglement, represents a resource for 
quantum information technology. Indeed, memory effects may mitigate 
the detrimental effects of the interaction with the external 
environment, such that quantum coherence may be preserved longer. 
This prompted efforts to precisely define the notion of NM for quantum 
processes, a task which have been pursued in different ways, with 
reference to different
mathematical properties of the quantum dynamical map. 
\par
In this paper, we stick with the BLP measure of non-Markovianity 
\cite{blp}, where Markovianity is thus seen as  an irreversible flow of 
information from the system to
the environment, while non-Markovianity allows for 
the information  to flow back  into the system.
Distinguishability between any two states can be defined using 
the trace distance  :
\begin{align} D[\rho_1 , \rho_2] = \frac{1}{2} \| \rho_1 - \rho_2 
\|_1. \end{align}
which provides a metric in the Hilbert space of physical states.
The trace distance takes values between 0 and 1 for indistinguishable  
and orthogonal states respectively.
Moreover, it is invariant under unitary transformations
$U$, $UU^\dag = U^\dag U= {\mathbb I}$, i.e. 
$D[U\rho_1 U^{\dagger} , U\rho_2 U^{\dagger}]=D[\rho_1 , 
\rho_2]$, and it is contractive, i.e. 
$D [\Phi \rho_1 , \Phi \rho_2 ] \le D [ \rho_1 ,  \rho_2 ]$,
for any completely positive and trace-preserving map $\Phi$.
If the distinguishability between states decreases, information flows 
out of the system into the environment, and the two expression above just
express the facts that information is preserved in closed systems, and 
that the maximum amount of information that can be recovered by the 
system cannot be larger than the amount that flowed out of it.
\par
In order to measure the degree of NM, one has to introduce the 
information flow
$\dot{D} = \frac{d}{d\tau} D[\rho_1 (\tau),  \rho_2(\tau)]   $
where $\rho_{12}(0)$ represent a pair of initial states.
A $\dot{D}> 0$ means a reversed flow of information.
For non-Markovian processes, there exist at least one pair of initial states and 
a temporal interval in which $\dot{D}$ is positive,
meaning that the trace distance between the initial states is increasing, i.e. information is flowing back.
The BLP measure of non-Markovianity quantifies the  growth in distinguishability, related to the total amount of information backflow, i.e.
\begin{align} 
\mathcal{N}\!(\tau) = \underset{\rho_{12}(0)}{\max} \,\underset{  \dot{D} > 0}{\int_0 ^{\tau}}\dot{D}[s,\rho_{12}(0)] ds .
\label{nonmar}
\end{align}
where the maximum is evaluated taking into account all pairs of initial 
states. The dynamics of the system is thus non-Markovian 
when $\mathcal{N} > 0$. 
%%%%%%%%%%%%%%%%
\subsection{Teleportation}\label{app2}
\begin{figure}[h!]
\centering
\includegraphics[width=0.95\columnwidth]{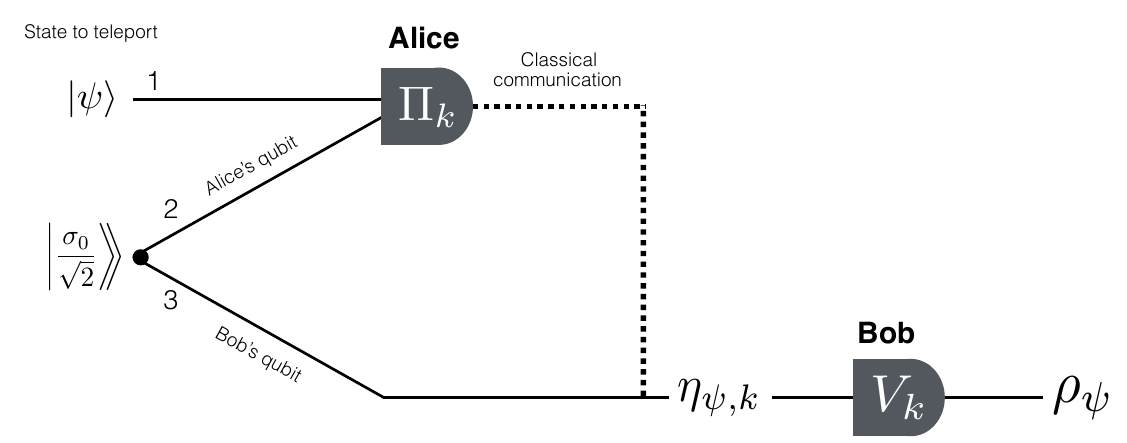}
\caption{Quantum teleportation protocol.
A and B share a maximally entangled state, then Alice performs a 
Bell measurement on qubits $1 \& 2$ and communicates which results 
she got to Bob, who performs a suitable unitary transformation on his
conditional state. The protocol works with neither Alice, 
nor Bob, knowing which state $|\psi\rangle$ is being teleported.}
\label{qtel}
\end{figure}

Quantum teleportation is a protocol where entanglement 
is exploited in order to transmit an unknown quantum 
state
$\ket{\psi}=\cos\frac{\theta}{2}\ket{0}+e^{i\phi}\sin\frac{\theta}{2}\ket{1}$ from Alice (A) to  Bob (B) without physically sending the qubit. In order for teleportation to work exactly,  A and B need to share a maximally
entangled state, e.g. $\ketket{\frac{\sigma_0}{\sqrt2}}$. The overall input state is the three-qubit state $\ket{\psi}_{\text{\scriptsize \!1}}\otimes\ketket{\frac{\sigma_0}{\sqrt2}}_{\text{\scriptsize \!23}}$. Then Alice performs a 
Bell measurement $\Pi_k=\ketket{\frac{\sigma_k}{\sqrt{2}}}_{\text{\scriptsize \!12}}\,{}_{\text{\scriptsize \!12}}\brabra{\frac{\sigma_k}{\sqrt{2}}}$ on her qubits 
(see the protocol scheme in Fig. \ref{qtel}).
 \par\noindent
The state of qubit $3$, conditional to Alice obtaining the result $k$, 
is given by  $\eta_{\psi,k} =  \sigma_k \,P_\psi\, \sigma_k$
where $P_\psi=\ket{\psi}\bra{\psi}$ is the projector of over the 
initial state of qubit $1$. In order to retrieve the input 
state, Bob should know which operation to perform in order 
to correct the effects of the reduction postulate. To this purpose, Alice 
communicates to Bob though a classical channel which results she has got.
Bob then implements a suitable unitary $V_k=\sigma_k$ on his qubit, 
obtaining the output state of the protocol 
$\rho_\psi=\sigma_k\eta_{\psi,k} \sigma_k=P_\psi$, $\forall k$, 
which is the input state that we wanted to teleport. 
We remind that the protocol works with neither Alice, 
nor Bob, knowing the teleported state.

\end{document}